\documentclass[aps,twocolumn,prd,showpacs,floatfix]{revtex4-1}

\usepackage{bm}
\usepackage{amsmath}
\usepackage{amsthm}
\usepackage{amssymb}
\usepackage{graphicx}

\newcommand{\eref}[1]{(\ref{#1})}
\newcommand{\eps }{\varepsilon }

\begin{document}

\title{Dense spectrum of resonances and particle capture in a near-black-hole
metric}
\author{V. V. Flambaum}
\affiliation{School of Physics, University of New South Wales, Sydney 2052,
Australia}
\author{G. H. Gossel}
\affiliation{School of Physics, University of New South Wales, Sydney 2052,
Australia}
\author{G. F. Gribakin}
\affiliation{School of Mathematics and Physics,
Queen's University, Belfast BT7 1NN, Northern Ireland, UK}

\date{\today}

\begin{abstract}
We show that a quantum scalar particle in the gravitational field of a massive
body of radius $R$ which slightly exceeds the Schwarzschild radius $r_s$,
possesses a dense spectrum of narrow resonances. Their lifetimes and density
tend to infinity in the limit $R \to r_s$. We determine the cross section
of the particle capture into these resonances and show that it is equal
to the absorption cross section for a Schwarzschild black hole.
Thus, a non-singular static metric acquires black-hole properties before the
actual formation of a black hole. 
\end{abstract}


\maketitle  

\section{Introduction}
The static Schwarzschild metric for a compact massive spherical body (i.e., a
black hole) has a coordinate singularity at the event horizon $r=r_s$, where
$r_s$ is the Schwarzschild radius. The absorption cross section by this object
is usually calculated by assuming a purely ingoing-wave boundary condition at
$r\to r_s$ \cite{Matzner,Starob,Unruh,Sanchez,Das,Crispino,Decanini}.
In particular,
for a massless scalar particle Unruh showed that $\sigma_{a} = 4\pi r_s^2$ at
zero energy \cite{Unruh}. More recently Kuchiev continued the wave function
analytically across the event horizon and found a nonzero gravitational
reflection coefficient $\mathcal{R}$ from $r=r_s$ \cite{RCalc,RadRef}, which
results in $\sigma_{a} =0$ at zero energy. Subsequently, the absorption cross
section was determined for an arbitrary reflection coefficient $\mathcal{R}$
which may, in principle, include gravitational, electromagnetic and other
interactions \cite{ScalarScattering}.

In this paper we consider the scattering problem for a massless scalar particle 
in a nonsingular metric of a massive body of radius $R>r_s$. We find that
in the limit $R \to r_s$, an increasingly dense spectrum of narrow
{\em resonances} emerges in the system. These quasistationary states exist in
the interior of the body of radius $R$, which resembles a ``resonant cavity''.
(Note that these resonances are different from the orbital resonaces which
exist outside a black hole, see, e.g., Refs.
\cite{Vishveshwara,Deruelle,Damour,Soffel,Pravica,Glampedakis,Grain}.)

For $R \to r_s$ both the resonance energy spacing $D$ and their width
$\gamma$ tend to zero, while their ratio remains finite,
e.g., $\gamma/D \simeq 2 \eps^{2}r_s^2/\pi$ for small energies $\eps $.
(We use units where $\hbar =c =1$.) This allows one to define the cross section
for particle capture into these long-lived states in the spirit of the optical
model \cite{LLV3}, by averaging over a small energy interval containing many
resonances. Note that this capture emerges in a purely potential scattering
problem, without any absorption introduced {\em a priori}. Somewhat
unexpectedly, the capture cross section turns out to be equal to the
cross section obtained by assuming total absorption at the event horizon
(i.e., for the reflection coefficient $\mathcal{R}=0$). In particular, in the
zero-energy limit our result coincides with Unruh's absorption cross
section for a black hole.

It is worth noting that the quantum scattering delay time associated with
the resonances, i.e., their lifetime $t =\hbar /\gamma $, is much longer than
the classical
gravitation dilation time. The resonance lifetime tends to infinity in the
limit $R\to r_s$, and the resonance capture becomes equivalent to absorption
(i.e., the particle does not come out during finite time). Therefore,
we observe a smooth, physical transition to the black-hole limit
with typical black-hole gravitational properties emerging for a nonsingular
static metric prior to the actual formation of the black hole.
We should add that in the case of finite-mass particles this picture is
complemented by a dense spectrum of the gravitationally {\em bound} states
located in the range $r<R$, see, e.g., Refs. \cite{Soffel,Gossel}.
Here too the spectrum becomes infinitely dense in the limit $R \to r_s$, and
the lowest level approaches $\eps =0$ (i.e., the binding energy is $-mc^2$).

We note that quantum effects (including the famous Hawking radiation) are
negligible for the star-mass black holes. Therefore, we do not aim to
consider real stars (see, e.g., \cite{comment2}). In the present
work we consider the theoretical question of how quantum effects manifest
themselves when a metric approaches the black hole metric.
\section{Scattering by static spherically symmetric body}
\label{sec:GeneralCase}

\subsection{Exterior solution}\label{subsec:ext}

The Klein-Gordon equation for a scalar particle of mass $m$ in a curved
space-time with the metric $g_{\mu \nu}$ is
\begin{equation}
\label{eq:KG}
\partial_\mu (\sqrt{-g} g^{\mu\nu}\partial_\nu \Psi)+\sqrt{-g} m^2\Psi =0.
\end{equation}
Outside a spherically symmetric, nonrotating body of mass $M$ and
radius $R$ the metric is given by the Schwarzschild solution
\begin{equation}\label{eq:ExteriorMetric}
ds^2 = \left(1-\frac{r_s}{r}\right)dt^2-
\left(1-\frac{r_s}{r}\right)^{-1}dr^2 -r^2 d\Omega^2 ,
\end{equation}
where $r_s = 2GM$, $G$ is the gravitational constant, and
$d\Omega ^2=d\theta ^2+\sin ^2\theta d\varphi^2 $.
For a particle of energy $\eps $ we seek solution of Eq.~(\ref{eq:KG}) in
the form $\Psi (x)=e^{-i\eps t}\psi (r)Y_{lm}(\theta ,\varphi )$. Considering
for simplicity the case of a massless particle in the $s$-wave ($l=0$), one
obtains the radial equation
\begin{equation}
\label{eq:ExteriorWave}
\psi''(r) + \left(\frac{1}{r-r_s}+\frac{1}{r}\right)\psi'(r)
+\frac{r^2 \eps ^2}{(r-r_s)^2}
\psi (r) = 0.
\end{equation}
If the radius of the body $R$ only slightly exceeds $r_s$, the first term
in brackets dominates for $r-r_s\ll r_s$, and the solution just
outside the body is the following linear combination of the incoming
and outgoing waves,
\begin{equation}\label{eq:R}
\psi \sim \exp \left[-ir_s\eps \ln \frac{r-r_s}{r_s}\right]+
\mathcal{R}\exp \left[ir_s\eps \ln \frac{r-r_s}{r_s}\right],
\end{equation}
where $\mathcal{R}$ is the reflection coefficient. It is determined either
by the boundary condition at $r=R$ (e.g., total absorption $\mathcal{R}=0$
imposed for a black hole \cite{Unruh}), or by matching the
solution with that at $r<R$ (e.g., analytically continuing to $r<r_s$
\cite{RCalc,RadRef}).

At large distances $r\gg r_s$ Eq.~(\ref{eq:ExteriorWave}) takes the form
of the nonrelativistic radial Shr\"odinger equation for a particle of unit
mass and momentum $\eps $ in the Coulomb-like potential $Z/r$ with
$Z=-\eps ^2r_s$. For $\eps r\gg 1$ its solution has the standard
form
\begin{equation}\label{eq:asymS}
\psi \propto r^{-1}\left(e^{-iz}-S e^{iz}\right),
\end{equation}
where $z=\eps r+\eps r_s\ln 2\eps r$, which defines the scattering
matrix $S$. To find $S$, one integrates Eq.~(\ref{eq:ExteriorWave})
outwards starting from one of the exponential solutions in Eq.~(\ref{eq:R}),
which gives a linear combination of ingoing and outgoing waves
at large $r$,
\begin{equation}\label{eq:albe}
\exp \left[-ir_s\eps \ln \frac{r-r_s}{r_s}\right]\longrightarrow \frac{r_s}{r}
\left[\alpha (\eps) e^{-iz}+\beta (\eps ) e^{iz}\right],
\end{equation}
where $|\alpha |^2-|\beta |^2=1$ due to flux conservation.
Comparison of Eqs.~(\ref{eq:R}), (\ref{eq:asymS}) and (\ref{eq:albe}) gives
the $S$ matrix as
\begin{equation}\label{eq:Sab}
S=-\frac{\beta +\alpha ^*\mathcal{R}}{\alpha +\beta ^*\mathcal{R}}.
\end{equation}

At low energies $\eps r_s\ll 1$, $\alpha $ and $\beta $ from
Eq.~(\ref{eq:albe}) can be found by matching the Coulomb solutions valid at
large distances with an intermediate-range solution obtained by
neglecting the last term in Eq.~(\ref{eq:ExteriorWave}) \cite{ScalarScattering},
\begin{eqnarray}\label{eq:alpha}
\alpha &=& \frac{i(1+\eps ^2r_s^2 C^2)}{2\eps r_sC}\exp(-i\delta _C),\\
\beta &=& -\frac{i(1-\eps ^2r_s^2 C^2)}{2\eps r_sC}\exp(i\delta _C),
\label{eq:beta}
\end{eqnarray}
where $C^2=2\pi\eps r_s /[1-\exp(-2\pi \eps r_s)]$ and
$\delta _C$ is the Coulomb phase shift \cite{LLV3}.

\subsection{Interior solution}\label{subsec:inter}

Consider a massive body with radius $R>r_s$ and interior metric
\begin{equation}\label{eq:Metric_ab}
ds^2 = a(r)dt^2-b(r)dr^2 -r^2 d\Omega^2 ,
\end{equation}
which matches that of Eq.~(\ref{eq:ExteriorMetric}) at the boundary:
$a(R)=\xi $ and $b(R)=\xi ^{-1}$, where $\xi =1-r_s/R$. 
For this metric the $s$-wave radial equation is
\begin{equation}\label{eq:KG_ab}
\frac{1}{r^2}\sqrt{\frac{a}{b}}\frac{d}{d r}\left( r^2\sqrt{\frac{a}{b}}
\frac{d \psi}{d r}\right)+\eps ^2\psi =0.
\end{equation}
Its analysis is particularly simple if we change the radial wavefunction
to $\phi =r\psi $, and the radial variable to the Regge-Wheeler ``tortoise''
coordinate $r^*$ defined by $d r^*=\sqrt{b/a}\,d r$. This gives the following
Schr\"odinger-like equation
\begin{equation}\label{eq:Sch}
\frac{d^2\phi }{d{r^*}^2}+\left[\eps ^2-\frac{1}{2r}\left(\frac{a}{b}\right) '
\right]\phi =0.
\end{equation}
The second term in brackets plays the role of an effective potential
for the motion in the $r^*$ coordinate. 

For a near-black-hole interior metric, $a(r)\to 0$ for $0\leq r\leq R$, as
the time slows down in the limit $\xi \to 0$. This means that the second term
in brackets in Eq.~(\ref{eq:Sch}) can be neglected for all except very small
energies, and the solution describes free motion in the tortoise coordinate.
The solution regular at the origin then is
\begin{equation}\label{eq:sol_int}
\phi \simeq \sin \eps r^* =\sin \left(\eps \int _0^r
\sqrt{\frac{b(r')}{a(r')}}\,dr'\right) .
\end{equation}
Matching this wave function to that of Eq.~(\ref{eq:ExteriorWave}) at $r=R$
yields the reflection coefficient
\begin{equation}\label{eq:TotalPhase}
\mathcal{R} = -\exp[2i \eps r_s B(\xi )],
\end{equation}
where 
\begin{equation}\label{eq:Bxi}
r_sB(\xi )=\int _0^R\sqrt{\frac{b(r)}{a(r)}}\,dr-r_s\ln \frac{R-r_s}{r_s}.
\end{equation}
Here the first term is due to the large phase acccumulated by the interior
solution. It increases much faster than the second one, and dominates for
$\xi \to 0$ where $B(\xi )\rightarrow \infty $. This large integral also gives
the classical time that a massless particle ($ds^2=0$) spends in the interior.
Its actual dependence on $\xi $ depends on the model used for the interior
metric (see Sec.~\ref{sec:Flo}). 

\subsection{Scattering matrix and resonances}\label{subsec:Smat}

For $\xi \ll 1$, the scattering matrix $S$ at low energies $\eps r_s\ll 1$ is
found explicitly using Eqs.~(\ref{eq:alpha}) and (\ref{eq:beta})
\cite{ScalarScattering},
\begin{equation} \label{eq:ScatMatrix}
S= \frac{1- \eps ^2r_s^2C^2 +(1+\eps ^2r_s^2C^2)\mathcal{R}}
{1+\eps ^2r_s^2C^2+ (1-\eps ^2r_s^2C^2)\mathcal{R}}e^{2i\delta_C}.
\end{equation}
Since the phase $2\eps r_s B(\xi )$ of $\mathcal{R}$ in
Eq. (\ref{eq:TotalPhase}) is large, one can show that the $S$-matrix has many
poles close to the real energy axis at $ \eps = \eps _n - i \gamma _n/2$
($n=1,\,2,\dots $). They correspond to resonances with positions and widths
\begin{eqnarray}\label{eq:eps_n}
 \eps _n &=&\frac{n\pi}{r_sB(\xi )}, \\ 
\gamma_n &=& \frac{2 \eps _n^2 r_sC^2}{B(\xi )}
=\frac{2\pi ^2 n^2C^2}{r_s[B(\xi )]^3}.\label{eq:Gam}
\end{eqnarray}
While both $\eps _n$ and $\gamma _n$ for a given $n$ depend on $\xi $ and
tend to zero in the limit $R\rightarrow r_s$, the {\em ratio} of the
width to the level spacing $D= \eps _{n+1}- \eps _n$,
\begin{equation}\label{eq:Ratio}
\gamma_n/D = 2 \eps ^{2}r_s^2 C^2/\pi ,
\end{equation}
is independent of $\xi $ for a given energy $\eps _n\approx \eps $.

Equation~(\ref{eq:Ratio}) shows that at low energies $ \eps r_s \ll 1$
the widths are much smaller than the resonance spacings. In this case
the cross section for the capture of the particle into these long-lived
states is described by the optical-model energy-averaged absorption cross
section \cite{LLV3}
\begin{equation} \label{eq:OpticalSigma}
\bar{\sigma}_{a}^{\rm opt}=\frac{2\pi^2}{\eps^2}\frac{\gamma_n}{D}.
\end{equation}
Substituting \eref{eq:Ratio} into \eref{eq:OpticalSigma} yields
\begin{equation}
\label{eq:FinalCrossSection}
\bar{\sigma}_{a}^{\rm opt}= 4\pi r_{s}^{2},
\end{equation}
which is equal to Unruh's low-energy absorption cross section for a black
hole \cite{Unruh}. Note that this result does not depend on $R$ or $B(\xi )$
for $\xi \ll 1$, that is, it is \textit{independent} of the particular interior
metric used.

\subsection{Absorption cross section}\label{subsec:absxs}

More generally, the optical-model absorption cross section is defined as
\begin{equation}\label{eq:siga}
\bar{\sigma}_{a}^{\rm opt}=\frac{\pi}{\eps ^2}(1-|\overline{S}|^2),
\end{equation}
where the $S$ matrix averaged over an energy interval contating many
resonances \cite{LLV3}. Without such averaging $|S|=1$, as long as
$|\mathcal{R}|=1$ in Eq.~(\ref{eq:Sab}), and the absorption cross section is
zero.

For the near-black-hole metric ($\xi \to 0$) the energy-averaging is
equivalent to averaging $S$ from Eq.~(\ref{eq:Sab}) over the large phase
$\Phi =\eps r_sB(\xi )$ of the reflection coefficient
$\mathcal{R}=-e^{2i\Phi }$. Since $\alpha $ and $\beta $ have a much slower
dependence on $\eps $ than $\Phi $, this gives
\begin{equation}\label{eq:betal}
\overline{S}=-\frac{1}{\pi }\int _0^\pi \frac{\beta -\alpha ^*e^{2i\Phi }}
{\alpha +\beta ^*e^{2i\Phi }}\,d\Phi =-\frac{\beta }{\alpha }.
\end{equation}
Hence, we see that averaging over the
large phase $\Phi $ is equivalent to setting $\mathcal{R}=0$ in
Eq.~(\ref{eq:ScatMatrix}), that is, to the assumption that there is no outgoing
wave in Eq.~(\ref{eq:R}) in the vicinity of the horizon.

At low energies $\eps r_s\ll 1$, the scattering matrix,
$S=\exp [2i(\delta_C+\delta )]$, and the corresponding short-range scatering
phase shift $\delta $ display
narrow resonances (see Sec.~\ref{sec:Flo}). The lifetime of these resonances
($\gamma _n^{-1}$) is related to the derivative of the phase shift
$d\delta /d\eps $ taken at the resonance, and is large for $\xi \rightarrow 0$.
More generally, in quantum mechanics, this derivative corresponds to the
expectation value of the {\it time delay} caused by trapping of the particle
in the potential well \cite{Smith}.

At higher energies, $\eps r_s\gtrsim 1$, the quantum resonances become broad,
and the motion becomes semiclassical. In this regime the phase shift $\delta $
is determined by the large (semiclassical) phase of the internal wavefunction in
Eq.~(\ref{eq:sol_int}). The corresponding quantum delay time is then given by
the classical dilation time $\int _0^R\sqrt{b(r)/a(r)}dr$, which also tends to
infinity for $\xi \rightarrow 0$.

The long trapping that one observes in the limit $R\rightarrow r_s$
at both low and high energies provides a physical explanation for the
no-reflection condition $\mathcal{R}=0$ that emerges as a result of averaging
of the scattering matrix in Eq.~(\ref{eq:betal}).

\section{Examples of interior metric}\label{sec:Flo}

In this section we illustrate our findings using particular models of the
interior metric. The standard Schwarzschild interior solution for a sphere
of constant density develops a pressure singularity when
$r_s = 8R/9$~\cite{Buchdahl}. This forbids the investigation of the
black-hole limit $R \rightarrow r_s $ in this model.
Examples of interior metric free from such singularity were proposed
by Florides~\cite{Flor74,comment2}
\begin{equation}
\label{eq:InteriorMetric}
a(r)=\frac{(1-r_s/R)^{3/2}}{\sqrt{1-r_sr^2/R^3}},\quad
b(r)=\left(1-\frac{r_s r^2 }{R^3}\right)^{-1},
\end{equation}
and Soffel \textit{et al.} \cite{Soffel}, for which
\begin{equation}\label{eq:metSoff}
a(r)= \left(1-\frac{r_s}{R}\right) 
\exp \left[-\frac{r_s (1-r^2/R^2)}{2R (1-r_s/R)}\right],
\end{equation}
and $b(r)$ is given by Eq. (\ref{eq:InteriorMetric}). Both metrics are valid
for $0\leq r\leq R$ for any $R>r_s$, and match the Schwarzschild solution
at $r=R$.

\subsection{Florides metric}\label{subsec:Flo}

For the Florides metric, using $a(r)$ and $b(r)$ from
Eq.~(\ref{eq:InteriorMetric}), we solve the second-order differential equation
(\ref{eq:KG_ab}) numerically with the boundary condition
$\psi (0)=1$, $\psi '(0)=0$ using {\em Mathematica} \cite{math}. This solution
provides a real boundary condition for the {\em exterior} wave function at
$r=R$. (We set $R=1$ in the numerical calculations).
Equation (\ref{eq:ExteriorWave}) is then integrated outwards to large
distances $r\gg r_s$. In this region Eq.~(\ref{eq:ExteriorWave})
takes the form of a nonrelativistic Shr\"odinger equation for a particle
with momentum $\eps $ and unit mass in the Coulomb potential with charge
$Z=-r_s\eps ^2$. Hence, we match the solution with the asymptotic form
\cite{LLV3}
\begin{equation}\label{eq:CoulombMatch}
\psi (r)\propto \sin [\eps r - (Z/\eps ) \ln 2 \eps r +\delta_C+ \delta ]
\end{equation}
where $\delta_C = \arg \Gamma (1+i Z/\eps ) $ is the Coulomb phase shift,
and determine the short-range phase shift $\delta $.

\begin{figure}[t!]
\begin{center}
\includegraphics*[width=0.48\textwidth]{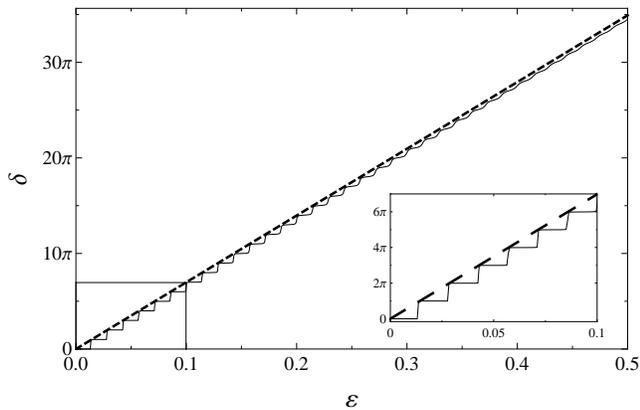}
\caption{Short-range phase shift $\delta $ as a function of energy obtained
numerically for $r_s=0.999 R$ (solid line). The dashed line shows the phase
accumulated by the wave function at $r\leq R$, $\Phi =\eps r_sB(\xi )$,
see Eqs. (\ref{eq:sol_int}), (\ref{eq:Bxi}), and (\ref{eq:B}).
\label{fig:TotalPhase}}
\end{center}
\end{figure}

The phase shift $\delta $ is due to the interior metric, and it carries
information about the behaviour of the wave function at $r\leq R$.
Unlike $\delta _C$ which is small and has a weak dependence on the energy,
$\delta $ depends strongly on the energy of the particle,
and is large for $r_s$ close to $R$ (i.e., for $\xi \ll 1$),
This is shown in Fig.~\ref{fig:TotalPhase} for $r_s=0.999 R$ (solid line).
The phase shift $\delta $ goes through many steps of the size $\pi$,
which correspond to the resonances described in Sec.~\ref{sec:GeneralCase}.
They occur approximately where the phase $\Phi = \eps r_sB(\xi )$
of the interior solution equals $n\pi $. For the Florides metric the phase
integral in Eq.~(\ref{eq:Bxi}) is
\begin{equation}\label{eq:A}
\int _0^R\sqrt{\frac{b}{a}}\,dr 
=\int _0^R\frac{\xi ^{-3/4}dr}{\sqrt[4]{1-r_sr^2/R^3}}
\simeq Ar_s\xi ^{-3/4},
\end{equation}
where $A=\sqrt{\pi }\Gamma (3/4)/[2\Gamma (5/4)]\approx 1.198 $, so that
\begin{equation}\label{eq:B}
B(\xi )=A\xi ^{-3/4}-\ln \xi .
\end{equation}
The corresponding phase $\Phi $ is shown in Fig.~\ref{fig:TotalPhase} by the
dashed line. Apart from the resonant steps, it matches closely the
short-range phase shift from the numerical calculation.

\begin{figure}[t!]
\begin{center}
\includegraphics[width=0.48\textwidth]{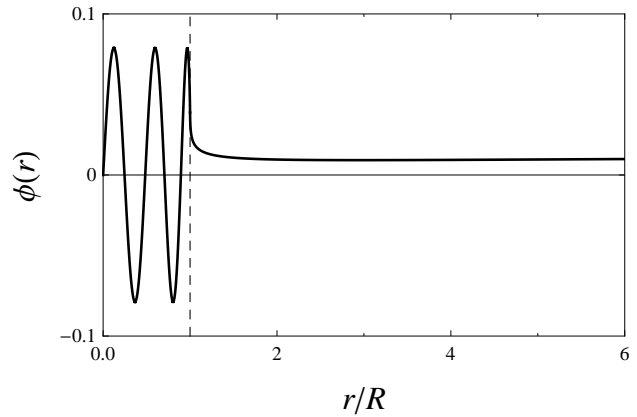}
\caption{Radial wave function $P(r)=r\psi (r)$ of
the fifth resonance, $\eps \approx 0.071$,
for $r_s = 0.999R$.}
\label{fig:Wavefunction}
\end{center}
\end{figure}

For the ``on-resonance'' energies corresponding to the midpoints of the steps
(where $d\delta /d\eps $ is largest) the magnitude of the wave function
$\phi (r)$ inside the body ($r<R$) is much greater than outside.
This is a signature of a quasistationary state of the trapped particle,
and is shown in Fig.~\ref{fig:Wavefunction}. 

The phase obtained using the Florides interior metric, Eq.~(\ref{eq:A}),
shows that the time dilation effect for the particle inside the massive body,
$d\delta /d\eps $,
increases as $(R-r_s)^{-3/4}$. The lifetimes of the resonances are in fact even
longer. As already mentioned in Sec. \ref{sec:GeneralCase}, the rapid variation
of the phase of $\mathcal{R}$ with energy gives rise to a sequence of resonant
poles in the $S$-matrix. The energies and widths of the resonances
are described by Eqs. (\ref{eq:eps_n}) and (\ref{eq:Gam}), respectively.
Numerically, they can be determined by fitting the ``steps''
in $\delta $ with $\arctan [2( \eps - \eps _n)/\gamma _n]$ \cite{LLV3}.

\begin{figure}[ht!]
\begin{center}
\includegraphics*[width=0.48\textwidth]{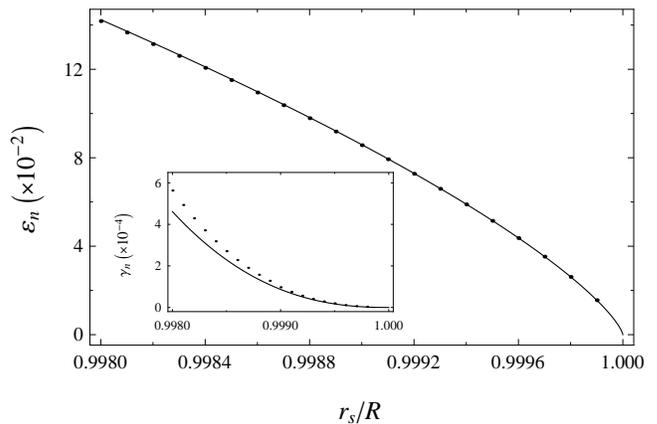}
\caption{Dependence of the energy $\eps _n$ (main plot) and width
$\gamma _n$ (inset) of the $n=6$ resonance on $r_s$ for the Florides
interior metric. Solid circles show the numerical values and solid lines
show the results obtained from Eqs. (\ref{eq:eps_n}) and (\ref{eq:Gam})
using $B(\xi )$ from Eq.~(\ref{eq:B}).
\label{fig:FloridesPositions}}
\end{center}
\end{figure}

Figure \ref{fig:FloridesPositions} shows that the dependence of the
resonance energy and width on $r_s$ obtained numerically and analytically
are in good agreement. In particular, we observe the rapid vanishing
of the width $\gamma _n \propto (R-r_s) ^{9/4}$ predicted by Eqs.~(\ref{eq:Gam})
and (\ref{eq:B}). Therefore, numerical calculation using the Florids metric
fully confirm the general analysis of the scattering problem
presented in Sec. \ref{sec:GeneralCase}.

\subsection{Soffel metric}\label{subsec:Sof}

\begin{figure}[t!]
\begin{center}
\includegraphics*[width=0.48\textwidth]{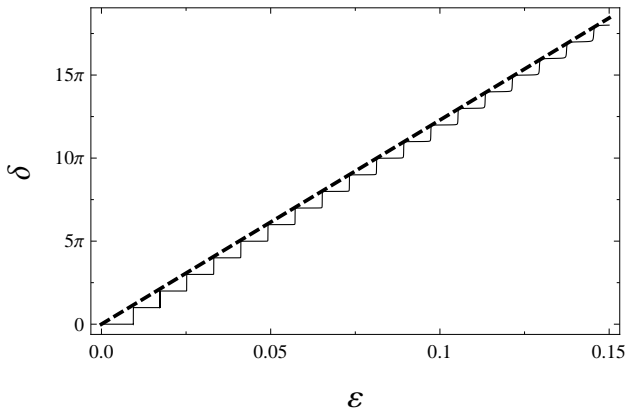}
\caption{Short-range phase $\delta $ obtained for the Soffel interior
metric for $r_s/R = 0.955$ (solid line), and the interior phase
$\Phi =\eps r_sB(\xi )$ (dashed line), with $B(\xi )$ given by
Eq.~(\ref{eq:B_Soff}).
\label{fig:SoffelPhase}}
\end{center}
\end{figure}

To verify our conclusions, we have also investigated the scattering problem
for $r_s$ close to $R$ using the Soffel interior metric, Eq.~(\ref{eq:metSoff}).
We do this numerically using {\em Mathematica}, as described 
in Sec.~\ref{subsec:Flo}. The corresponding short-range scattering
phase shift $\delta $ is shown in Fig.~\ref{fig:SoffelPhase} for
$r_s=0.955 R$. It displays resonance steps similar to those in
Fig.~\ref{fig:TotalPhase}.

\begin{figure}[ht!]
\begin{center}
\includegraphics*[width=0.46\textwidth]{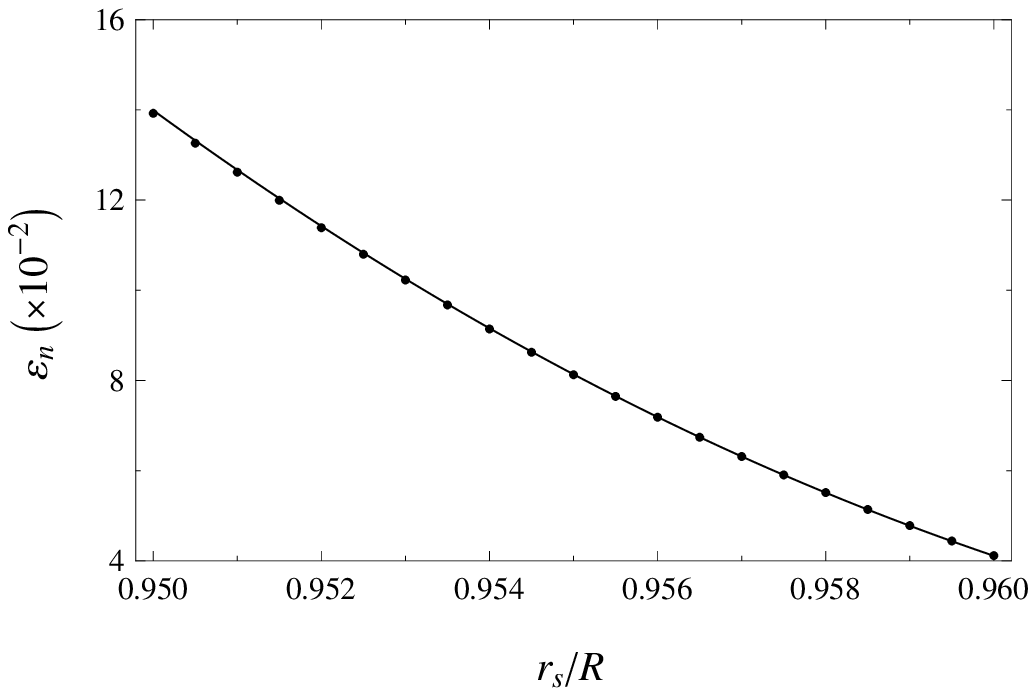}
\vspace{-12pt}
\caption{Energy of the $n=10$ resonance obtained for the Soffel
interior metric as a function of $r_s$: solid circles -- numerical values;
solid line -- analytical result, Eqs.~(\ref{eq:eps_n}) and (\ref{eq:B_Soff}).
\label{fig:SoffelPositions}}
\end{center}
\end{figure}

For the Soffel metric the leading contribution to the interior wave function
phase $\Phi $ is
\begin{equation}\label{eq:Soff_ph}
\int _0^R\sqrt{\frac{b}{a}}\,dr =
\int _0^R\frac{\exp \left[r_s (1-r^2/R^2)/(4R\xi) \right]}
{\sqrt{\xi (1-r_sr^2/R^3)}}\,dr,
\end{equation}
which for $\xi \ll 1$ gives [cf. Eq.~(\ref{eq:B})]
\begin{equation}\label{eq:B_Soff}
B(\xi )\simeq \sqrt{\frac{\pi }{1-3\xi }}\exp \left(\frac{1-\xi }{4\xi}\right)
-\ln \xi .
\end{equation}
This expression shows that for the Soffel
metric the short-range phase $\delta $, the resonance level density and
their lifetimes increase exponentially for $r_s\to R$, i.e., for
$\xi \rightarrow 0$. This explains why in the Soffel metric the onset 
of the resonant scattering picture similar to that seen in the 
Florides metric occurs earlier, i.e., at smaller values of $r_s/R $. We see
that the dependence of $B(\xi )$, $\Phi $ and the
short-range phase $\delta $ on $\xi $ is not inversal. However, the dependence
of the short-range phase and resonances on the large parameter $B(\xi )$
[$B(\xi )\rightarrow \infty $ for $\xi \rightarrow 0$],
Eqs.~(\ref{eq:ScatMatrix})--(\ref{eq:Gam}), is universal.

\begin{figure}[t!]
\begin{center}
\includegraphics*[width=0.46\textwidth]{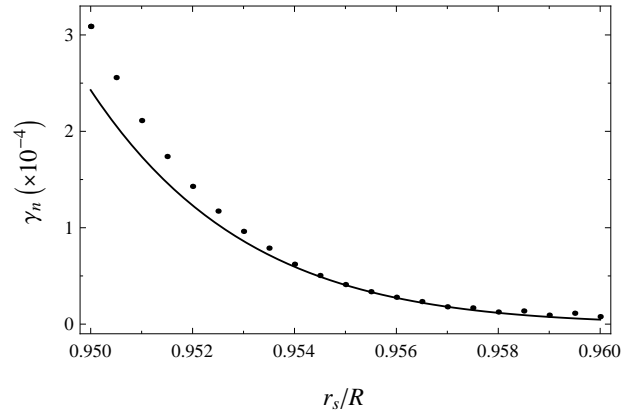}
\vspace{-12pt}
\caption{Width of the $n=10$ resonance obtained for the Soffel
interior metric as a function of $r_s$: solid circles -- numerical values;
solid line -- analytical result, Eqs.~(\ref{eq:eps_n}) and (\ref{eq:B_Soff}).
\label{fig:SoffelWidths}}
\end{center}
\end{figure}

Figure \ref{fig:SoffelPositions} shows the dependence of the resonance energy
$\eps _n$ on $r_s/R$ for $n=10$. It also shows that the numerical values
obtained by fitting the short-range phase with the resonant profile
$\arctan [2( \eps - \eps _n)/\gamma _n]$ (shown by circles), are in good
agreement with those obtained from Eqs.~(\ref{eq:eps_n}) and (\ref{eq:B_Soff})
(solid line).

As in the case of the Florides metric, the resonance widths $\gamma _n$
display a much faster descrease than the energies $\eps _n$ for
$r_s\rightarrow R$. This is shown for $n=10$ in Fig.~\ref{fig:SoffelWidths}.
The strong dependence of $\gamma _n$ on $r_s/R$ is explained by the rapid
(exponential) increase of $B(\xi )$ with decreasing $\xi $, see 
Eqs.~(\ref{eq:Gam}) and (\ref{eq:B_Soff}).

Hence, we see that the picture of resonant scattering for a metric close to
the black-hole limit, is independent of the particular interior metric used.


\section{Conclusions}
The problem of scattering of low-energy scalar particles from a massive
static spherical body has been considered. We have shown that as the
black-hole metric limit is approached, a dense spectrum of long-lived resonances
emerges in the problem. Long-time-delay trapping of the particles
in these resonances gives rise to effective absorption in a purely potential
scattering problem. We are grateful to the anonymous referee for the important
observation that the existence of the narrow resonances may be linked to the
``no-hair''theorem, as the particle must be trapped inside in the limit
$R=r_s$.

This shows that the absorption boundary condition ($\mathcal{R}=0$)
emerges naturally in the limit $R\to r_s$, as a result of particle
capture into the dense spectrum of long-lived resonances.

\end{document}